\documentclass[trackchanges,twocolumn,resetfootnote]{aastex7}
\usepackage{amsmath, amsthm, amssymb, amsfonts}
\usepackage{graphicx}
\usepackage{physics}
\usepackage{natbib}
\usepackage[dvipsnames]{xcolor}
\usepackage{float}
\usepackage{mathtools}
\usepackage{booktabs}
\usepackage{multirow}

\newcommand{\BB}{\mathbf{B}}
\newcommand{\EE}{\mathbf{E}}
\newcommand{\vv}{\mathbf{v}}
\newcommand{\uu}{\mathbf{u}}

\newcommand{\JJ}{\mathbf{J}}
\newcommand{\urad}{u_\mathrm{rad}}
\newcommand{\Grad}{\gamma_\mathrm{rad}}
\newcommand{\Trad}{\Theta_\mathrm{rad}}
\newcommand{\brec}{\beta_\mathrm{rec}}
\newcommand{\me}{m_\mathrm{e}}
\newcommand{\Ne}{n_\mathrm{e}}
\newcommand{\Te}{\Theta_\mathrm{e}}

\newcommand{\Le}{\lambda_\mathrm{e}}
\newcommand{\fe}{f_\mathrm{e}}
\newcommand{\fr}{f_\mathrm{r}}
\newcommand{\st}{\sigma_\mathrm{T}}
\newcommand{\FIC}{\mathbf{F}_\mathrm{IC}}
\newcommand{\etaeff}{\eta_\mathrm{eff}}
\newcommand{\bg}{\beta_\Gamma}
\newcommand{\vA}{v_\mathrm{A}}
\newcommand{\Ek}{E_\mathrm{kin}}
\newcommand{\Eb}{E_\mathrm{bulk}}
\newcommand{\tic}{t_\mathrm{IC}}
\newcommand{\tadv}{t_\mathrm{adv}}
\newcommand{\tx}{t_\mathrm{D}}

\defcitealias{Golan2015}{GL15}
\defcitealias{Moran2025}{M25}
\defcitealias{Ripperda2026}{R26}
\defcitealias{Daughton2006}{W. Daughton et al. 2006}
\defcitealias{Sironi2016}{L. Sironi et al. 2016}

\begin{document}
%TC:ignore
\title{Making Reconnection Cooler: A relativistic resistive magnetohydrodynamics model for radiative collisionless reconnection}

\author[0000-0003-0147-9562,gname="Alicia Marie",sname="Savelli"]{Alicia M. Savelli}
\affiliation{Canadian Institute for Theoretical Astrophysics, University of Toronto, 60 St. George Street, Toronto, ON M5S 3H8, Canada}
\affiliation{David A. Dunlap Department of Astronomy \& Astrophysics, University of Toronto, 50 St. George Street, Toronto, ON M5S 3H4, Canada}
\affiliation{Dunlap Institute for Astronomy \& Astrophysics, University of Toronto, 50 St. George Street, Toronto, ON M5S 3H4, Canada}
\email[show]{alicia.savelli@utoronto.ca}

\author[0000-0002-5408-3046,gname="Daniel",sname="Groselj"]{Daniel Gro\v selj}
\affiliation{Centre for mathematical Plasma Astrophysics, Department of Mathematics, KU Leuven, Celestijnenlaan 200B, B-3001 Leuven, Belgium}
\email[]{daniel.groselj@kuleuven.be}

\author[0000-0002-7301-3908,gname="Bart",sname="Ripperda"]{Bart Ripperda}
\affiliation{Canadian Institute for Theoretical Astrophysics, University of Toronto, 60 St. George Street, Toronto, ON M5S 3H8, Canada}
\affiliation{David A. Dunlap Department of Astronomy \& Astrophysics, University of Toronto, 50 St. George Street, Toronto, ON M5S 3H4, Canada}
\affiliation{Department of Physics, University of Toronto, 60 St. George St, Toronto, ON M5S 1A7, Canada}
\email{ripperda@cita.utoronto.ca}
\affiliation{Perimeter Institute for Theoretical Physics, 31 Caroline St. North, Waterloo, ON N2L 2Y5, Canada}

\author[0000-0001-7572-4060,gname="Amir",sname="Levinson"]{Amir Levinson}
\affiliation{School of Physics and Astronomy, Tel Aviv University, Tel Aviv 69978, Israel}
\email{levinson@tauex.tau.ac.il}

\author[0000-0001-7801-0362,gname="Alexander",sname="Philippov"]{Alexander Philippov}
\affiliation{Department of Physics, University of Maryland, College Park, MD 20742, USA}
\affiliation{Joint Space-Science Institute, University of Maryland, College Park, MD 20742, USA}
\affiliation{Physics Department, Stanford University, 382 Via Pueblo Mall, Stanford, CA 94305, USA}
\affiliation{Kavli Institute for Particle Astrophysics and Cosmology, Department of Physics, Stanford University, Stanford, CA 94305, USA}
\email{sashaph@umd.edu}

\author[0000-0002-8659-3729,gname="Norm",sname="Murray"]{Norman Murray}
\affiliation{Canadian Institute for Theoretical Astrophysics, University of Toronto, 60 St. George Street, Toronto, ON M5S 3H8, Canada}
\affiliation{David A. Dunlap Department of Astronomy \& Astrophysics, University of Toronto, 50 St. George Street, Toronto, ON M5S 3H4, Canada}
\affiliation{Department of Physics, University of Toronto, 60 St. George St, Toronto, ON M5S 1A7, Canada}
\email{murray@cita.utoronto.ca}

\correspondingauthor{Alicia M. Savelli}

\begin{abstract}
Relativistic magnetic reconnection around astrophysical compact objects is a likely site for particle energization and multiwavelength emission.  Kinetic theory has substantially modelled such collisionless radiative relativistic reconnection using a radiative drag force countering the acceleration of the particles.  Here, we systematically derive an energy-momentum sink term in relativistic resistive magnetohydrodynamics (RRMHD), informed by kinetic theory and coupled with a charge-starved resistivity prescription which reproduces the fast reconnection rate seen in kinetic calculations, to model radiative losses.  We show that our radiative loss term accurately captures important macroscopic properties -- including the temperature in the reconnection layer and the bulk motion energy distribution of the out-flowing plasma -- for inverse Compton cooling applied to two-dimensional RRMHD simulations of a reconnection layer and benchmarked to kinetic simulations.  Our results hold in various cooling regimes with a one-to-one mapping to kinetic theory.  Our radiative model lays a foundation for future RRMHD simulations of plasma in the relativistic radiative regime, such as in reconnection sites in a local turbulent flow or globally in the magnetospheres of black holes and neutron stars.
% \ams{175 words}

\end{abstract}
%TC:endignore
\keywords{\uat{Plasma astrophysics}{1261} --- \uat{High energy astrophysics}{739} --- \uat{Magnetohydrodynamics}{1964} --- \uat{Magnetohydrodynamical simulations}{1966} --- \uat{Compact radiation sources}{289} --- \uat{Special relativity}{1551}
}

\section{Introduction\label{sec:intro}}

Magnetic reconnection is a canonical dissipation channel in astrophysical sources. One of the longest- and best-studied examples is the solar corona, where reconnection drives flares and contributes to the heating of the surrounding plasma \citep[e.g.,][]{Antolin2021,Cheng2023,Patel2025}. More recently, interest has expanded towards more distant and extreme sources, such as accreting black holes \citep[e.g.,][]{Parfrey2019,Sironi2020,Sridhar2021,Mehlhaff2021,Ripperda2022,Hakobyan2023b,Galishnikova2023,Gupta2024,Hakobyan2025,Sridhar2025,Mehlhaff2026}, neutron-star magnetospheres and their winds \citep[e.g.,][]{Cerutti2017, Philippov2019,Cerutti2020,  Hakobyan2023a,Mahlmann2024}, astrophysical jets \citep[e.g.,][]{Levinson2006,Petropoulou2016,Sironi2021}, and gamma ray bursts \citep[e.g.,][]{Lyubarsky2010, McKinney2012, Granot2016}. Reconnection in high-energy environments commonly occurs in the relativistic regime, which differs from more conventional (e.g., solar) plasmas, such that the magnetic energy density upstream of the reconnection layer far exceeds the plasma rest mass energy density (i.e., a high upstream magnetization parameter $\sigma_0\equiv B_0^2/4\pi\rho_0c^2\gg1$, where $B_0$ is the strength of the upstream magnetic field, $\rho_0$ is the upstream mass density, and $c$ is the speed of light), leading to efficient nonthermal particle acceleration \citep{Sironi2014,Guo2014,Werner2016}. 

Reconnection in high-energy sources is also frequently accompanied by rapid radiative cooling, which competes against the particle acceleration.  This cooling introduces new effects that alter the physics of the reconnection layer and its emission signature. Such effects include cooling of the particles on timescales faster than the dynamic dissipation time, radiative drag on the bulk motions, and -- in the most extreme cases -- reprocessing of the plasma composition via electron-positron pair creation and annihilation \citep{Hakobyan2019,Schoeffler2023,Mehlhaff2024}.  Therefore, in order to predict emission from high-energy sources powered by reconnection, there is a need to include, at some level, both radiative effects as well as relativistic kinetic physics.

One notable example of such a high-energy source undergoing rapid cooling is the magnetospheres of neutron stars, where reconnection can power multiwavelength emission \citep{Philippov2018,Philippov2019,Lyubarsky2020,Mahlmann2022,Hakobyan2023a,Parsons2026}.  Another prominent candidate for relativistic radiative reconnection is the black hole ``corona" -- hot plasma of moderate optical depth \citep{Yuan2004,Fabian2015} near the event horizon of luminous accreting black holes.  Here, radiative reconnection has been proposed as a mechanism for plasma energization \citep{Beloborodov2017,Sironi2020,Sridhar2021}, powering hard X-ray emission \citep{Zdziarski2004} believed to be sourced from the Comptonization of soft seed photons by relativistically hot electrons of temperature $k_\mathrm{B}T_\mathrm{e}\sim100$keV \citep{Eardley1975,Shapiro1976}, where $k_\mathrm{B}$ is the Boltzmann constant, as well as by the bulk plasma flow \citep{Beloborodov2017, Sridhar2025}.

% Magnetic reconnection in astrophysical systems often occurs in the radiative regime in the relativistic limit, and is relevant to emission observed from black hole flares \ams{hayk}, neutron star magnetosphere flares \ams{jens, lyubarsky},  pulsars beyond the light cylinder \ams{sashaP, alex chen, ashley}, relativistic jets \ams{giannios pertropoulou bart \#2}, and gamma ray bursts \ams{help}.  Non-relativistic reconnection has been known to play an important role in the heating of the solar corona \ams{help}, and its relativistic counterpart is often assumed to aid in the heating of black hole ``coronae" \ams{andrei} -- hot plasma of moderate optical depth \ams{Daniel \#5-6} nearby the event horizon of luminous accreting black holes.  

% In particular, luminous X-ray binaries (e.g., Cyg X-1 \ams{Daniel \#3}) radiate in a ``soft-state" of quasi-thermal emission generated from an optically thick accretion disk, and a ``hard-state" of non-thermal emission emanating from this so-called corona \ams{Z\&G}.  The emission mechanism for these hard X-rays is believed to be the Comptonization of soft seed photons by relativistically hot electrons of temperature $k_\mathrm{B}T_\mathrm{e}\sim100$keV \ams{Z2000 cites  Eardley, Lightman, \& Shapiro (1975) and Shapiro, Lightman, \& Eardley (1976)}, where $k_\mathrm{B}$ is the Boltzmann constant, as well as by the bulk plasma flow \ams{Andrei}.

Reconnection in astrophysical plasmas also typically occurs in the collisionless regime (i.e., the mean free path of a particle is much larger than the system size).  Kinetic particle-in-cell (PIC) simulations are required to accurately model the collisionless microsphysics of the plasma, but these simulations are computationally limited to a local reconnection layer and are able to capture neither their formation nor their interaction with the accretion flow of a global astrophysical system.  This separation of scales is especially relevant in the case of black hole coronae, where there has been considerable effort towards understanding the plasma-radiation interaction from first principles with the corona structured as both a single large-scale reconnection layer 
\citep{Sironi2020,Sridhar2021,Mehlhaff2021} and as a turbulent plasma driving multiple smaller current sheets \citep{Groselj2024,Nattila2024,Groselj2026}.  Recent work has also shown large-scale reconnecting current sheets forming within the putative coronal region in global general relativistic magnetohydrodynamic (GRMHD) simulations of an accreting black hole \citep{Liska2022,Sridhar2025}; however, the GRMHD framework lacks essential microphysical components of collisionless reconnection, such as a fast reconnection rate or physically-informed models for coronal plasma-radiation interaction and cooling, which could significantly impact global accretion.  There exist subgrid models for thermal IC cooling -- neglecting bulk Comptonization -- in ideal GRMHD simulations \citep{Kinch2020}, but these have never been applied directly to magnetic reconnection, utilized resistive magnetohydrodynamics, or been compared to kinetic results. 

In this letter, we develop, from first principles, a new subgrid model for collisionless relativistic radiative reconnection in resistive relativistic magnetohydrodynamics (RRMHD).  The model is  based on a set of systematically derived sink terms in the RRMHD energy and momentum equations that capture losses from both IC scattering and bulk Comptonization. We also utilize a charge-starved effective resistivity prescription \citep{Moran2025,Ripperda2026} to mimic the fast reconnection rate of collisionless plasmas. As a first step, we develop our model in the optically thin regime in an isotropic background radiation field. 
%In this paper \ams{letter?}, we develop, from first principles, a subgrid model for relativistic radiative reconnection in charge-starved resistive relativistic magnetohydrodynamics (RRMHD) in the optically-thin-to-moderate regime.  We simulate a two-dimensional (2D) local-box reconnection layer in pair $e^\pm$ plasma RRMHD and apply a sink term to model both thermal and bulk Comptonization.  
We validate our results against first-principles PIC simulations with the full microphysics of collisionless relativistic radiative reconnection, and find that our RRMHD subgrid model is able to reproduce macroscopic properties -- including the fast reconnection rate, the temperature in the reconnection layer, and the bulk motions of the outflow -- from PIC.  This confirmation of kinetic theory sets the stage for physically-informed global simulations of radiation-dominated systems with emission powered through magnetic reconnection.

\section{Model for radiative cooling\label{sec:cooling}} 
\subsection{Radiative drag force on particles}
To model IC cooling in kinetic theory, particles are considered to be travelling through a photon bath described by an isotropic radiation energy density $\urad$%\footnote{The assumption of isotropy is valid within our optically thin model. In the case of moderate optical depth, some anisotropies may result from the scattering of photos by the large-scale bulk motions \ams{Daniel \#51}.}
, causing the particles to scatter off the background photons.  Momentum losses due to IC scattering are calculated by subtracting a radiative drag force $\FIC$ from the Lorentz force experienced by the particles \citep{Landau1971}:
\begin{equation}
    \FIC = -\frac{4}{3}\st\urad\gamma^2\boldsymbol{\beta},
    \label{eq:FIC-1}
\end{equation}
where $\st$ is the Thomson cross section,  $\boldsymbol{\beta}=\mathbf{v}/c$ is the particle three-velocity $\mathbf{v}$ in units of $c$, and $\gamma=1/\sqrt{1-\beta^2}$ is the particle Lorentz factor.

The magnitude of $\FIC$ can be expressed in terms of a characteristic Lorentz factor $\Grad$ at which the radiative drag is perfectly balanced against the acceleration supplied by the electric field $E_\mathrm{rec}=\brec B_0$ in the reconnection layer 
\begin{equation}
    e\brec B_0=\frac{4}{3}\st\urad\Grad^2,
    \label{eq:force-balance}
\end{equation}
where $\brec$ is the reconnection rate, $e$ is the particle charge, and we have taken $\beta\to1$ assuming that particles accelerate in the diffusion region \citep{Uzdensky2016,Sironi2020}.  Equation~\eqref{eq:force-balance} gives a convenient parametrization for $\urad$
% \begin{equation}
%     \urad=\frac{3e\brec B_0}{4\st\Grad^2}
%     \label{eq:urad-gamma}
% \end{equation}
in terms of $\Grad$, 
% \begin{equation}
%     \FIC=-e\brec B_0\left(\frac{\Ge}{\Grad}\right)^2\boldsymbol{\beta},
%     \label{eq:FIC-2}
% \end{equation}
so that the strength of the cooling can now be controlled by the characteristic Lorentz factor such that smaller $\Grad$ corresponds to larger momentum losses and thus stronger cooling, and vice versa.  

This parametrization allows us to rewrite the drag force in terms of dimensionless parameters, the characteristic Lorentz factor, and a length scale, which together fully determine the physical regime of the radiative reconnection.  The most meaningful length scale is the half-thickness of the reconnection layer, estimated to be the hot Larmor radius $\Delta\sim r_\mathrm{L}=\gamma \me c^2/eB_0\sim B_0/4\pi n_{\mathrm{e},0} e$, using that $\gamma\sim\sigma_0$ in the diffusion regions around the X-points \citep{Kirk2003,Werner2016,Comisso2019,Chernoglazov2023}, and where $\me$ is the particle mass and $n_{\mathrm{e},0}$ is the particle number density in the upstream (which we take to be the same for both species).  Then, $\FIC$ can be written as
\begin{equation}
    \FIC=-\frac{\me c^2\brec\sigma_0}{\Delta}\left(\frac{\gamma}{\Grad}\right)^2\boldsymbol{\beta}.
    \label{eq:FIC-2}
\end{equation}

\subsection{Energy-momentum sink in fluid theory}
% \ams{
% \begin{itemize}
%     \item amir's eqns
%     \item can use F balance if cs
%     \item effective resistivity
%     \item theta
%     \item eqns with m/e
%     \item skin depth
%     \item Delta and rL
%     \item eqns in terms of Delta
% \end{itemize}
% }

The radiative losses in a fluid description of the plasma can be modelled as energy and momentum losses in the form of a sink term applied to the RRMHD equations.  Then, the evolution of the plasma is described by the continuity equation
\begin{align}
    \nabla_\nu\left(\rho u^\nu\right) &= 0,
    \label{eqn:continuity}\\
    \intertext{the equations for energy and momentum conservation,}
    \nabla_\nu T^{\mu\nu} &= \frac{1}{c}S^\mu,
    \label{eqn:energy-mom-cons}\\
    \intertext{and Maxwell's equations}
    \nabla_\nu {^*\!F}^{\mu\nu} &= 0,\\
    \nabla_\nu F^{\mu\nu} &= \frac{4\pi}{c}J^\mu.
\end{align}
where $\rho$ is the fluid mass density; $u^\mu$ is the fluid four-velocity; $T^{\mu}_{\nu}$ is the total stress-energy tensor, including both the fluid and electromagnetic parts; $S^\mu$ is the volumetric cooling rate sink term; $F^{\mu\nu}$ is the Faraday tensor and ${^*\!F}^{\mu\nu}$ is its Hodge dual; and 
\begin{equation}
    J^\mu=\left\{\varrho c,\varrho \vv+\frac{\Gamma}{\eta}\left[\EE+\frac{\vv\times\BB}{c}-\frac{\left(\EE\cdot\vv\right)\vv}{c^2}\right]\right\}
\end{equation}
is the resistive four-current, with charge density $\varrho$, Lorentz factor $\Gamma$, resistivity $\eta$, and velocity, electric, and magnetic field vectors $\vv$, $\EE$, and $\BB$, respectively.  This system of equations is closed by the ideal gas equation of state $p=(\hat{\gamma}-1)\rho\epsilon$, relating the pressure $p$ and mass density $\rho$ via the specific internal energy density $\epsilon$ and the adiabatic index $\hat{\gamma}$, which we set to 4/3.

The IC losses have both a component due to individual relativistic electrons upscattering the background photons, as well as a bulk Comptonization component due to photons getting upscattered instead by bulk motions of the plasma (i.e., the outflow in a reconnection layer) \citep{Beloborodov2017}, both of which require a counterpart in a fluid theory.  In adapting a fluid representation of IC cooling from kinetic theory, it is necessary to average over the particle distribution function so that individual particle quantities may be written in their analogous fluid form.  In particular, averaging over the individual particle motions translates to a temperature and a bulk velocity.  To model IC losses, the energy-momentum sink term $S^\mu$ should thus have a dependence on both temperature (capturing losses due to electron IC scattering) and bulk velocity (capturing losses due to bulk Comptonization).  

Following \citet{Golan2015}, the IC radiative sink term $S^\mu$ as a volumetric cooling rate for an isotropic radiation field is given by 
\begin{align}
    S^0 &=-\frac{4}{3}\frac{\st c}{\me}\,\urad\rho_\mathrm{e}\Gamma^2
    \left[
        \bg^2 + 12\Te^2
        \left(
            1+\bg^2
        \right)
    \right];\label{eq:S0-1}\\
    S^i & = -\frac{4}{3}\frac{\st c}{\me}\,\urad\rho_\mathrm{e}\Gamma^2\bg^i
    \left[
        1 + 4\Te^2
        \left(
            5+\bg^2
        \right)
    \right],\label{eq:Si-1}
\end{align}
where $\rho_\mathrm{e}$ is the electron mass density, $\Gamma=1/\sqrt{1-\bg^2}$ is the Lorentz factor of the bulk motions, $\bg^i$ is the bulk three-velocity $v_\Gamma^i$ in units of $c$, and $\Te=k_\mathrm{b}T_\mathrm{e}/\me c^2$ is the dimensionless electron temperature.  A full derivation of Equations~\eqref{eq:S0-1}-\eqref{eq:Si-1} can be found in Appendix~\ref{app:sink}\footnote{Note that in the limit of no bulk motions (i.e., $\Gamma^2\to1,\ \bg^2\to0$), $S^0$ recovers the volumetric cooling rate derived by \citet{Kinch2020} (their Equation~(8)) in the relativistic limit, $S^0=(4/3)\st\Ne c\urad(12\Te^2)$.  }.

%To parametrize the cooling strength in RRMHD in a similar fashion as in the particle drag force using the force balance from Equation~\eqref{eq:force-balance}, we introduce an effective resistivity to model sub-grid kinetic length scales such that the electric field in the reconnection layer $E_\mathrm{rec}$ matches that in kinetic theory.  
In order to quantitatively compare the RRMHD radiative losses $S^\mu$ to the kinetic radiative drag $\FIC$ (Equation \eqref{eq:FIC-2}), we use the same force-balance parametrization from Equation~\eqref{eq:force-balance}.  We therefore require that the electric field in the reconnection layer $E_\mathrm{rec}$ matches that in kinetic theory, and thus we introduce an effective resistivity to model the subgrid kinetic length scales.  \citet{Moran2025} (hereafter \citetalias{Moran2025}) proposes such a charge-starved resistivity: if the diffusion regions surrounding the X-points are sufficiently charge-starved, then the corresponding current density there is given by $|\JJ|\sim2\Ne ec$, where $\Ne$ is the particle number density and the 2 arises from counter-streaming species. 

These charge-starved regions can be represented in RRMHD by a localized, effective resistivity $\etaeff=|\EE^*|/|\JJ|\sim\brec B_0/2\Ne e c$, where $\EE^*=\Gamma\left[\EE+(\vv\times\BB)/c - (\EE\cdot\vv)\vv/c^2\right]$ \citep{Blackman1993,Komissarov2007}.  Such a prescription has been shown to reproduce the structure and magnitude of $\EE^*$, and hence the reconnection rate, from PIC simulations in the non-radiative regime \citep[hereafter \citepalias{Ripperda2026}]{Ripperda2026}. The charge-starved resistivity introduces an effective cold upstream skin depth $\Le=\sqrt{\me^2c^2/4\pi\rho_0e^2}$, parameterized in the numerical prescription by a mass-to-charge ratio $\me/e$ and upstream mass density $\rho_0$, such that the typical sheet half-thickness can be written concisely in RRMHD as a product of fundamental plasma parameters $\Delta\sim\sqrt{\sigma_0}\Le$.  %This relation can be used to eliminate the mass-to-charge ratio in Equations~\eqref{eq:S0-2} and \eqref{eq:Si-2}, such that the radiative sink term $S^\mu$ can be written as

To write the force-balance parametrization (Equation~\eqref{eq:force-balance}) fully in fluid form, we average over the thermal electron distribution function, such that $\left\langle\gamma^2\right\rangle=12\Te^2$ in the relativistic limit \citep{Rybicki1979}.  In the case of a pair plasma\footnote{If the plasma also consists of ions, an additional electron temperature equation may be included \citep{Ressler2015}.}, we can take $\rho_\mathrm{e}=\rho/2$ and $\Te=\Theta$, where $\rho$ and $\Theta$ are the RRMHD mass density and dimensionless temperature, respectively. Then, $\urad$ can be written in terms of a characteristic $\Trad$ as the force-balance in Equation~\eqref{eq:force-balance} becomes $e\brec B_0=16\st\urad\Trad^2$. Reparametrizing Equations~\eqref{eq:S0-1} and \eqref{eq:Si-1} in terms of $\Delta$ and $\Trad$, $S^\mu$ becomes
% \begin{align}
% S^0 &
% \begin{multlined}[t]
%     =-\frac{1}{12}\frac{\rho e}{m}c\brec B_0\Gamma^2
%     \left(
%     \frac{\Theta}{\Trad}
%     \right)^2\times\\
%     \left[
%         \left(
%             \frac{\bg}{\Theta}
%         \right)^2 + 12
%         \left(
%             1+\bg^2
%         \right)
%     \right];
% \end{multlined}
%     \label{eq:S0-2}\\
%     S^i & 
%     \begin{multlined}[t]
%         = -\frac{1}{12}\frac{\rho e}{m}c\brec B_0\Gamma^2\bg^i
%     \left(
%     \frac{\Theta}{\Trad}
%     \right)^2\times\\
%     \left[
%         \frac{1}{\Theta^2} + 4
%         \left(
%             5+\bg^2
%         \right)
%     \right].
%     \end{multlined}\label{eq:Si-2}
% \end{align}

\begin{align}
        S^0 &
    \cramped{
    =-\frac{c^3 \brec \sigma_0}{24\Delta} 
    \rho\Gamma^2
    {\left(\!
    \frac{\Theta}{\Trad}
    \!\right)\!}^2
    \left[\!
        {\left(\!
            \frac{\bg}{\Theta}
        \!\right)\!}^2 + 12
        \left(
            1+\bg^2
        \!\right)
    \!\right]
    }
    ;\\
    S^i & 
    \cramped{
    =-\frac{c^3 \brec \sigma_0}{24\Delta} 
    \rho\Gamma^2\bg^i
    {\left(\!
    \frac{\Theta}{\Trad}
    \!\right)\!}^2
    \left[\!
        \frac{1}{\Theta^2} + 4
        \left(
            5+\bg^2
        \!\right)
    \right]
    }
    .
\end{align}

% \begin{align}
% S^0 &
% \begin{multlined}[t]
%     =-\frac{1}{12}\frac{B_0^2}{4\pi\Delta}\brec\frac{\rho}{\rho_0}c\Gamma^2 
%     \left(
%     \frac{\Theta}{\Trad}
%     \right)^2\times\\
%     \left[
%         \left(
%             \frac{\bg}{\Theta}
%         \right)^2 + 12
%         \left(
%             1+\bg^2
%         \right)
%     \right];
% \end{multlined}
%     \label{eq:S0-Trad}\\
%     S^i & 
%     \begin{multlined}[t]
%         = -\frac{1}{12}\frac{B_0^2}{4\pi\Delta}\brec\frac{\rho}{\rho_0}c\Gamma^2 \bg^i
%     \left(
%     \frac{\Theta}{\Trad}
%     \right)^2\times\\
%     \left[
%         \frac{1}{\Theta^2} + 4
%         \left(
%             5+\bg^2
%         \right)
%     \right].
%     \end{multlined}\label{eq:Si-Trad}
% \end{align}

Thus, in the relativistic regime, where  the current density in diffusion regions approaches charge starvation \citepalias{Moran2025,Ripperda2026}, the radiative loss term for RRMHD recovers the parametrization of the radiative drag from kinetic theory, allowing a one-to-one mapping of cooling strengths from PIC $\to$ RRMHD via $\Grad^2\to12\Trad^2$.  %We therefore use $\Grad$ and $\Trad$ interchangeably throughout the remainder of this work.  

\subsection{Radiative cooling regimes\label{sec:cooling regimes}}
When defining the strength of the cooling in models of radiative reconnection, there are two characteristic length scales of interest: (1) the typical size of a diffusion region, relevant to the distance over which individual particles are accelerated and cooled; and (2) the typical length of the reconnection layer, relevant to bulk Comptonization.  To identify the cooling regimes associated with each of these length scales, one must compare the respective crossing time with the typical cooling time \citep{Sironi2020}:
\begin{equation}
    \tic = \frac{\me c\gamma\boldsymbol{\beta}}{\mathbf{F}_\mathrm{IC}} = \frac{\Delta\Grad^2}{\brec\sigma_0 c\gamma},\label{eq:tIC}
\end{equation}
using Equation~\eqref{eq:FIC-2}.

First considering cooling on the scale of a diffusion region, the associated light-crossing time is given by $\tx=\ell_\mathrm{D}/c$, where $\ell_\mathrm{D}\sim\Delta/\brec$ is the length of the diffusion region measured parallel to the reconnection layer, and we estimate the aspect ratio of the diffusion region as $\Delta/\ell_\mathrm{D}\sim\brec$ \citepalias{Ripperda2026}.  Then the condition for strong cooling on this scale is given by $\tic/\tx\lesssim1$, or $\Grad^2\lesssim\sigma_0^2$, using that electrons are typically accelerated up to a Lorentz factor $\gamma\sim\sigma_0$ in the diffusion regions in the absence of cooling.  Conversely, weak cooling on this scale occurs when $\tic/\tx\gg1$, or $\Grad^2\gg\sigma_0^2$.

In the case of cooling along the length of the reconnection layer, strong cooling occurs when the cooling time scale is much shorter than the timescale for plasma advection $\tic/\tadv\ll1$, where $\tadv\sim L/\vA\sim L/c$, $L$ is the half-length of the sheet, and $\vA$ is the Alfv\'en speed.  In terms of the characteristic Lorentz factor, the condition for strong cooling is then $\Grad^2\ll\brec L \sigma_0/\Delta$, where we have assumed $\gamma\sim1$.  Likewise, the cooling on this scale is weak when $\tic/\tadv\gg1$, or $\Grad^2\gg\brec L \sigma_0/\Delta$.

We therefore study four separate cooling regimes to cover our entire parameter space: (1) \textbf{strong cooling} $\left(\tic\lesssim\tx\right)$, (2) \textbf{moderate cooling} $\left(\tx\ll\tic\ll\tadv\right)$, (3) \textbf{weak cooling} $\left(\tic\gg \tadv\right)$, and (4) \textbf{no cooling }$\left(\tic\to\infty\right)$.  We calculate values for $\Grad$ for each of the four cooling regimes using simulation parameter values $\sigma_0=15$, $L\approx3000\Le$ with roughly 5 cells per $\Le$, and fiducial $\brec=0.1$.  Our corresponding choices for $\Grad$ and its mapping to $\Trad$ in each regime are summarized in Table~\ref{tab:cooling}.  For the remainder of this paper, we shall list cooling strenghts using the PIC $\Grad$ only, for brevity, and readers may refer to Table~\ref{tab:cooling} for the corresponding RRMHD $\Trad$.
\begin{deluxetable}{ccc}
\tablecaption{The characteristic Lorentz factors $\Grad$ and dimensionless temperatures $\Trad$ corresponding to each of the four cooling regimes defined in Section~\ref{sec:cooling regimes} chosen for the simulation runs with magnetization $\sigma_0=15$ and half-sheet length $L\approx3000\Le$. The analogous cooling strengths between PIC and RRMHD are calculated using the mapping $\Grad^2\to12\Trad^2$. \label{tab:cooling}}
\tablehead{
\colhead{Regime} & \colhead{$\Grad$ (PIC)} & \colhead{$\Trad$ (RRMHD)}
}
\startdata
\multirow{4}{*}{\shortstack{\textbf{Strong cooling}\\$\tic\lesssim\tx$}} & 10 & 2.9 \\
 & 12 & 3.5 \\
 & 14 & 4.0 \\
 & 15 & 4.3 \\
\hline
\multirow{4}{*}{\shortstack{\textbf{Moderate cooling}\\$\tx\ll\tic\ll\tadv$}} & 17 & 4.9 \\
 & 20 & 5.8 \\
 %& 25 & 7.2 \\
 & 32 & 9.2 \\
\hline
\addlinespace[4pt]
\multirow{1}{*}{\shortstack{\textbf{Weak cooling}\\$\tadv\gg\tic\ll\tadv$}}  & 60 & 17.3
\\\addlinespace[8pt]
\hline
\textbf{No cooling} & $\infty$ & $\infty$ \\
\enddata
\end{deluxetable}

These regimes map onto distinct astrophysical sources depending on the role IC scattering plays in each. In black hole coronae, where IC dominates the radiative losses \citep{Nattila2024,Groselj2024}, particles leave the diffusion regions before radiating \citep{Beloborodov2017,Sironi2020}, yet the IC cooling time is far shorter than the advection time along the layer, placing coronal current sheets in the moderate regime; in this regime, Compton drag on the plasmoids converts most of the dissipated energy into radiation through bulk Comptonization \citep{Sironi2020,Sridhar2021}. Gamma-ray burst jets, where the radiation energy density exceeds the magnetic energy density, are expected to be qualitatively similar, while the IC-dominated jets of flat-spectrum radio quasars \citep{Ghisellini2009} lie near the boundary between moderate and weak cooling. In contrast, strong cooling at the diffusion regions occurs in the magnetospheric current sheets of young pulsars and M87* through synchrotron losses \citep{Chernoglazov2023,Hakobyan2023b}, where IC scattering is dynamically negligible yet can produce the observed TeV emission \citep{Hakobyan2023b}. 

\section{Simulation setup \label{sec:sims}}
We apply our radiative model to RRMHD simulations of a single reconnection layer with a charge-starved resistivity prescription \citepalias{Moran2025,Ripperda2026} in Minkowski spacetime using the \texttt{Black Hole Accretion Code} \citep[\texttt{BHAC,}][]{BHAC1,BHAC2,BHAC3}, and benchmark the results against first-principles radiative PIC simulations of pair plasma run with \texttt{TRISTAN-MP v2} \citep{tristan1,tristan_v2}.  

Both simulation setups feature a 2D Harris current sheet \citep{Harris1962} in Cartesian ($x,y$) coordinates with the sheet parallel to the $x$-axis.  We initialize the PIC and RRMHD simulations such that the initial conditions are as similar as possible. The magnetic field is initialized by the reconnecting magnetic field $B_x=B_0\tanh(y/d)$ -- reversing sign at $y=0$ over a thickness $d$ of order a few tens of skin depths -- and by the uniform guide field aligned with the electric current $B_z=B_g=0.5B_0$.  We include a moderate guide field to represent reconnection in turbulent systems, where idealized anti-parallel configurations (i.e., $B_g \approx 0$) are rare \citep[e.g.,][]{Zhdankin2013,Chernoglazov2021,Nurisso2023,Groselj2024,Ha2025,Groselj2026,Davis2026}, and to provide magnetic pressure to support plasmoids against collapsing due to radiative losses.  The electric and velocity fields are initialized to 0.  The initial mass density profile is given by $\rho=\rho_0+\rho_\mathrm{s}/\cosh^2(y/d)$ with an over-density in the sheet $\rho_\mathrm{s}=3\rho_0$ relative to the upstream density set by the upstream magnetization $\sigma_0=15$, chosen to be in the relativistic regime.  In RRMHD, the pressure profile is then $p=p_0+B_0^2/(2\cosh^2\left(y/d)\right)$ such that the plasma pressure inside the sheet balances the magnetic pressure, and where the upstream pressure is $p_0 = 10^{-2}\rho_0c^2$ such that the upstream dimensionless temperature in both PIC and RRMHD is set to $\Theta_0=10^{-2}$.  To trigger reconnection, the pressure balance is disturbed at the centre of the current sheet by a small perturbation, generating an X-point.  

Our numerical domain has a length of $2L$ and a comparable effective resolution of $N_x\times N_y=32,768^2$ in RRMHD (using a base resolution of $N_x\times N_y=256^2$ with eight levels of adaptive mesh refinement applied to the current sheet region of the domain) and $N_x\times N_y=30,720\times30,000$ in PIC.  In both RRMHD and PIC, we set a half-box size of $L\approx3000\Le$, such that we resolve roughly $5$ cells per skin depth. 
%We choose a mass-to-charge ratio of $m/e=8\times10^{-5}$ in MHD, corresponding to a sheet half-thickness $\Delta_0\approx0.001L$, and a box size of $L\approx3000\Le$, %($L_x\approx3277\Le$ in RRMHD and $L_x=3072\Le$ in PIC) 
%such that we resolve $\sim5$ cells per skin depth in both RRMHD and PIC.  
We confirm in Appendix~\ref{app:resolution} that the effective skin depth scale is well-resolved and hence so is the diffusion region.  The boundaries in the outflow direction are initially periodic and switch to absorbing after one light-crossing time $t_c=L/c$ \citepalias{Daughton2006,Sironi2016,Ripperda2026}, while the boundaries parallel to the sheet employ ``plasma injectors" in PIC and absorbing boundaries (effectively replenishing the initial magnetic field and density) in RRMHD.  The radiation prescription also turns on after one $t_c$, where we start solving the radiative source term $S^\mu$ (Equations~\eqref{eq:S0-1}-\eqref{eq:Si-1}) explicitly after the resistive source term \citep{BHAC3} in RRMHD, and $\FIC$ (Equation~\eqref{eq:FIC-2}) during the particle push in PIC \citep{Tamburini2010,Vranic2016,Zou2025}.  We set a threshold of $\gamma>\gamma_\mathrm{thresh}=1.1$ and $\Gamma>\Gamma_\mathrm{thresh}=1.02$ for which to apply the radiation prescription in PIC and RRMHD, respectively\footnote{Note that the thresholds differ as $\gamma_\mathrm{thresh}$ in PIC refers to the particle Lorentz factor -- which includes the higher-energy particles in the Maxwell-J\"uttner distribution, while $\Gamma_\mathrm{thresh}$ in RRMHD is the bulk Lorentz factor -- which is generally lower than the maximum particle Lorentz factor $\gamma$.}, in order to avoid cooling the upstream plasma.

\section{Results}
% \ams{
% Plan:
% \begin{itemize}
%     \item vrec: doesn't change w cooling and why
%     \item Particle energy spectrum: how to get for mhd, why mhd and pic don't match for no cooling, asymptote at $\Grad$, doesn't match exactly but don't need it to
%     \item bulk spectrum: very good match, peaks at inflow/outflow
%     \item cooling moves spectra left
%     \item big takeaway: mhd captures macroscipic pic -- vrec and bulk motions (but not plasmoids or energy spectrum but that's okay bc not necessary for global)
% \end{itemize}
% }
Here, we demonstrate that our RRMHD cooling model can faithfully reproduce -- both qualitatively and quantitatively -- the essential macroscopic results of radiative PIC simulations, performed in matching conditions for a given relative cooling strength.  For our analysis, we select one representative cooling strength per regime for which to compare: $\Grad\in\{\infty,60,32,15\}$ for no, weak, moderate, and strong cooling, respectively.  We focus on macroscopic features important for future large-scale GRMHD simulations, rather than microscopic physics only probeable through kinetic theory.

\begin{figure*}[htb!]
    \centering
    \includegraphics[width=\linewidth]{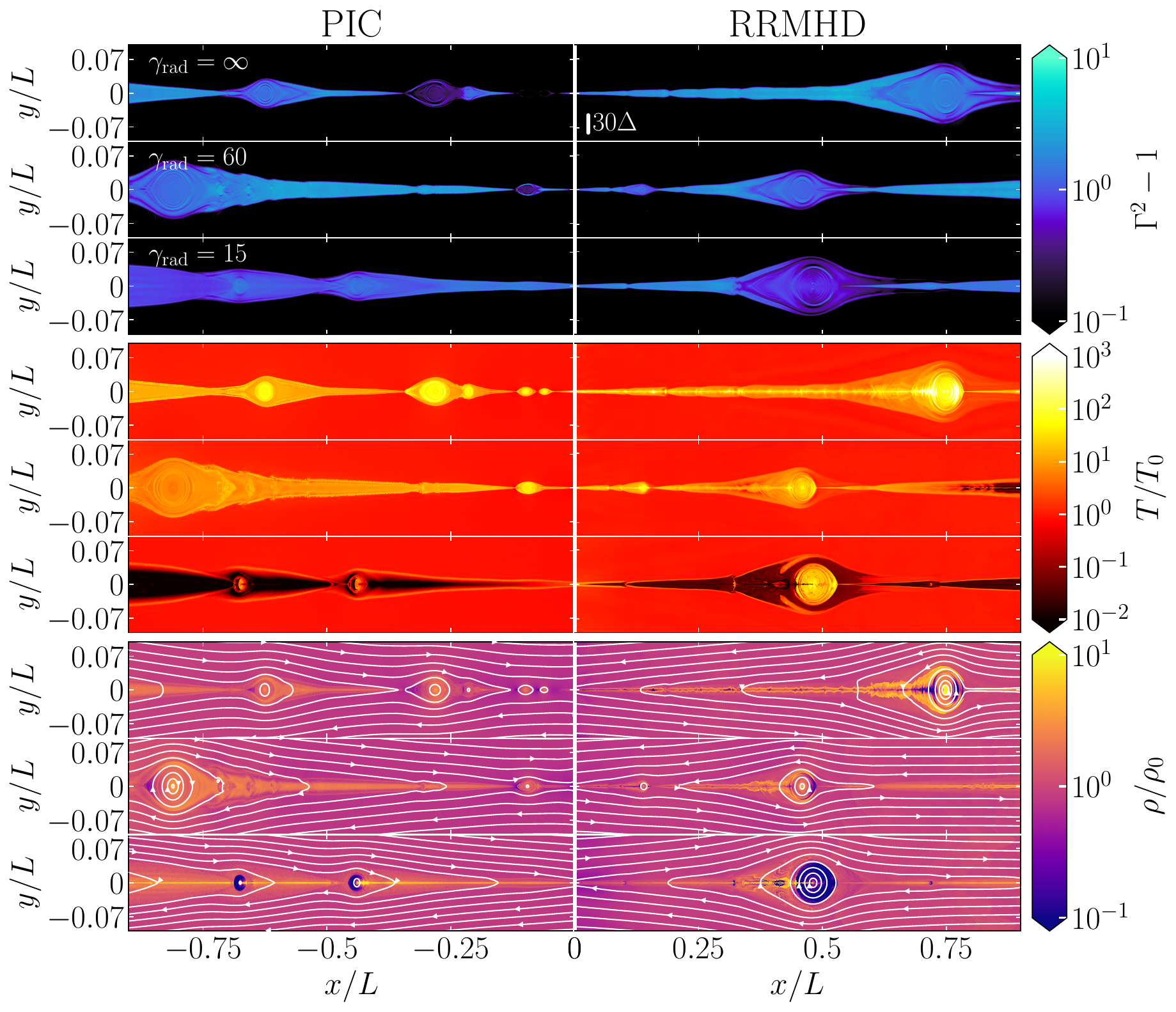}
    \caption{Comparison of the reconnection layer structure in PIC (\textbf{left}) and MHD (\textbf{right}) for a range in cooling strengths. Three blocks each display a different quantity: bulk speed $\Gamma^2-1$ (\textbf{top}), temperature normalized by the upstream $T/T_0$ \textbf{(middle)}, and density normalized by the upstream $\rho/\rho_0$ overlaid with white contours of the out-of-plane vector potential pointing in the direction of the in-plane magnetic field (\textbf{bottom}).  A scale-bar indicating the relative size of $30\Delta$ is provided in the top-right panel, indicating the order of the size of the sheet thicknesses.  Each block shows the given quantity in three cooling regimes: no cooling ($\Grad=\infty$; \textbf{top}), weak cooling ($\Grad=60$; \textbf{middle}), and strong cooling ($\Grad=15$; \textbf{bottom}).  With stronger cooling, the temperature in the sheet is noticeably reduced, and the sides of plasmoids opposing the out-flowing motion become less dense.  The effect of cooling on the bulk motions is secondary to temperature, but the sheet and plasmoids both show lower $\Gamma$ with stronger cooling.}
    \label{fig:sheet}
\end{figure*}

We first examine the structure of the reconnection layer across varying cooling strengths in both the RRMHD and PIC models in Figure~\ref{fig:sheet}.
For all cooling regimes, RRMHD accurately reproduces the overall morphology of the reconnection layer as well as the typical magnitudes of the bulk speed, plasma temperature, and density in the sheet with respect to the PIC result.  Here, $T=\me p/k_\mathrm{B}\rho$ is the plasma temperature, and ``bulk speeds" (i.e., the square of the bulk four-velocity, $\Gamma^2\bg^2=\Gamma^2-1$) are found from the bulk Lorentz factor via the bulk three-velocities $\Gamma\equiv\Gamma(v_\Gamma)$; in PIC $v_\Gamma$ is calculated from the magnitude of the current density $|\JJ|=\varrho |\vv_\Gamma|$, where $\varrho$ is the charge density.  The effect of stronger cooling is clearest in temperature -- the temperature of the sheet drastically drops with decreasing $\Grad$ -- with a secondary effect decreasing the bulk motion speeds for decreasing $\Grad$ -- most noticeable in the plasmoids and at the far edge of the outflow.  There is also a decrease in density on the plasmoid faces opposing the direction of the bulk flow with stronger cooling \citep{Sironi2020}.

\begin{figure}[htb!]
    \centering
    \includegraphics[width=\linewidth]{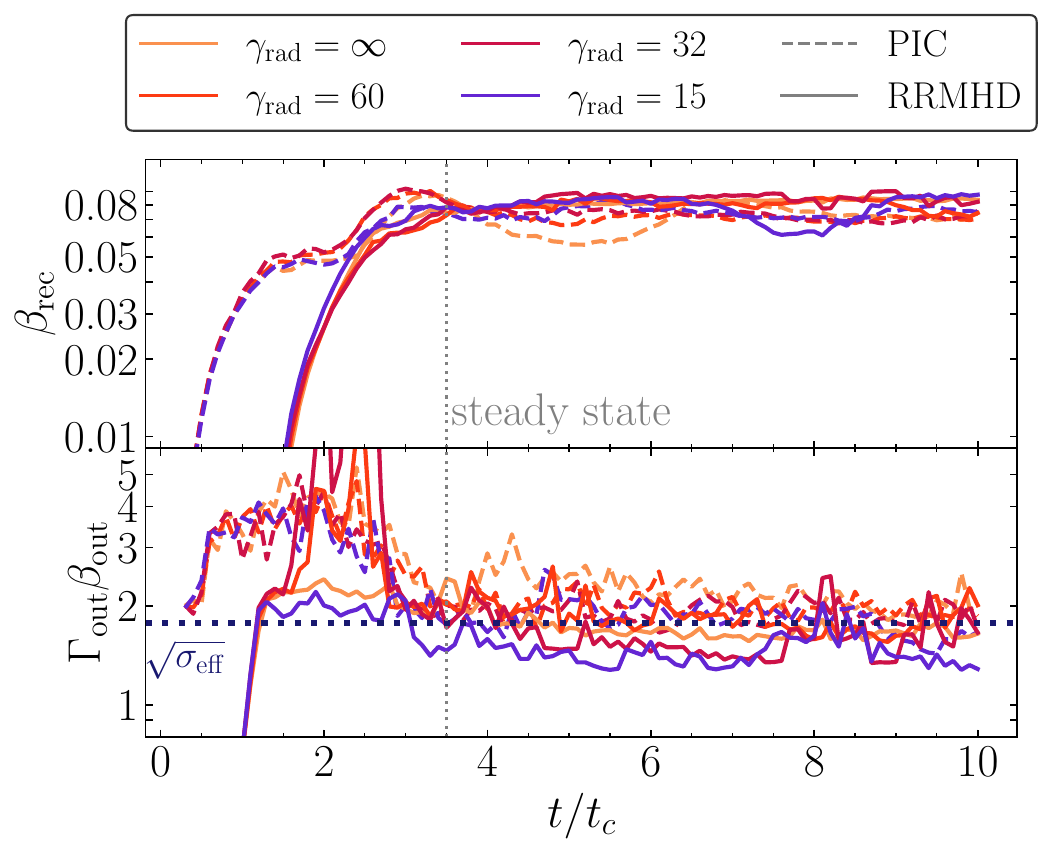}
    \caption{\textbf{Top:} The time-evolution of the reconnection rate $\brec$ in each of the four cooling regimes (colours), showing good agreement between PIC (dashed lines) and MHD (solid lines).  When time-averaged over the steady state ($t\geq3.5t_c$; indicated by grey dotted line), we find a rate of $\brec=0.08$ across all simulations shown.  \textbf{Bottom:} The time-evolution of the four-velocity in the outflow direction $\Gamma_\mathrm{out}\beta_\mathrm{out}$.  Both MHD and PIC converge to the analytic Alfv\'enic outflow $\Gamma_\mathrm{A}\beta_\mathrm{A}\sim\sqrt{\sigma_\mathrm{eff}}$ (dotted navy line).
    The strength of the cooling has negligible effect on $\brec$ or $\Gamma_\mathrm{out}\beta_\mathrm{out}$.}
    \label{fig:vrec}
\end{figure}

We next consider characteristics of the inflow and outflow speeds with respect to the reconnection layer. The inflow speed is found from averaging over slices in the velocity perpendicular to the current sheet, $\beta_\mathrm{in}=\langle\beta_y\rangle$; while the outflow speed is taken to be the maximum speed parallel to the current sheet, $\beta_\mathrm{out}=\max\left(|\beta_x|\right)$.  With the charge-starved resistivity model, RRMHD is able to recover the kinetic reconnection rate $\langle\brec\rangle=\langle\beta_\mathrm{in}/\beta_\mathrm{out}\rangle\sim0.08$ when time-averaged over the steady state past the initial transient ($t>3.5t_c$) and over all four cooling regimes, as shown in the top panel of Figure~\ref{fig:vrec}.
%The average reconnection rate in PIC alone is $0.072\pm0.005$, and in RRMHD alone is $0.081\pm0.005$, matching within one standard deviation.  
Note that the reconnection rate is mediated by the guide field and thus is slightly lower than the typically quoted $\sim0.1$ rate for zero-guide reconnection \citep{Gupta2024}.  The steady-state bulk outflow speed is shown in the bottom panel of Figure~\ref{fig:vrec}, and is consistent between PIC and RRMHD for all cooling strengths, well-matching the analytic Alfv\'{e}nic outfow 
%$\Gamma_\mathrm{A}\beta_\mathrm{A}\sim\sqrt{\sigma_\mathrm{eff}}=B_0^2/\left\{4\pi\rho_0c^2\left[1+\hat{\gamma}p_0/\left(\left(\hat{\gamma}-1\right)\rho_0c^2\right)\right]+B_g^2\right\}$ 
$\Gamma_\mathrm{A}\beta_\mathrm{A}=\sqrt{B_0^2\sigma_0/(B_0^2+B_g^2\sigma_0)}\sim\sqrt{\sigma_\mathrm{eff}}$ \citep{Lyubarsky2005,Takamoto2016,Sironi2025}, where $\beta_\mathrm{A}$ is the Alfv\'en speed in units of $c$ and $\Gamma_\mathrm{A}$ is the corresponding Lorentz factor.  The strength of the cooling has negligible effect on the reconnection rate or outflow speed, in agreement with previous studies \citep{Sironi2020}.  

\begin{figure}[h]
    \centering
    \includegraphics[width=\linewidth]{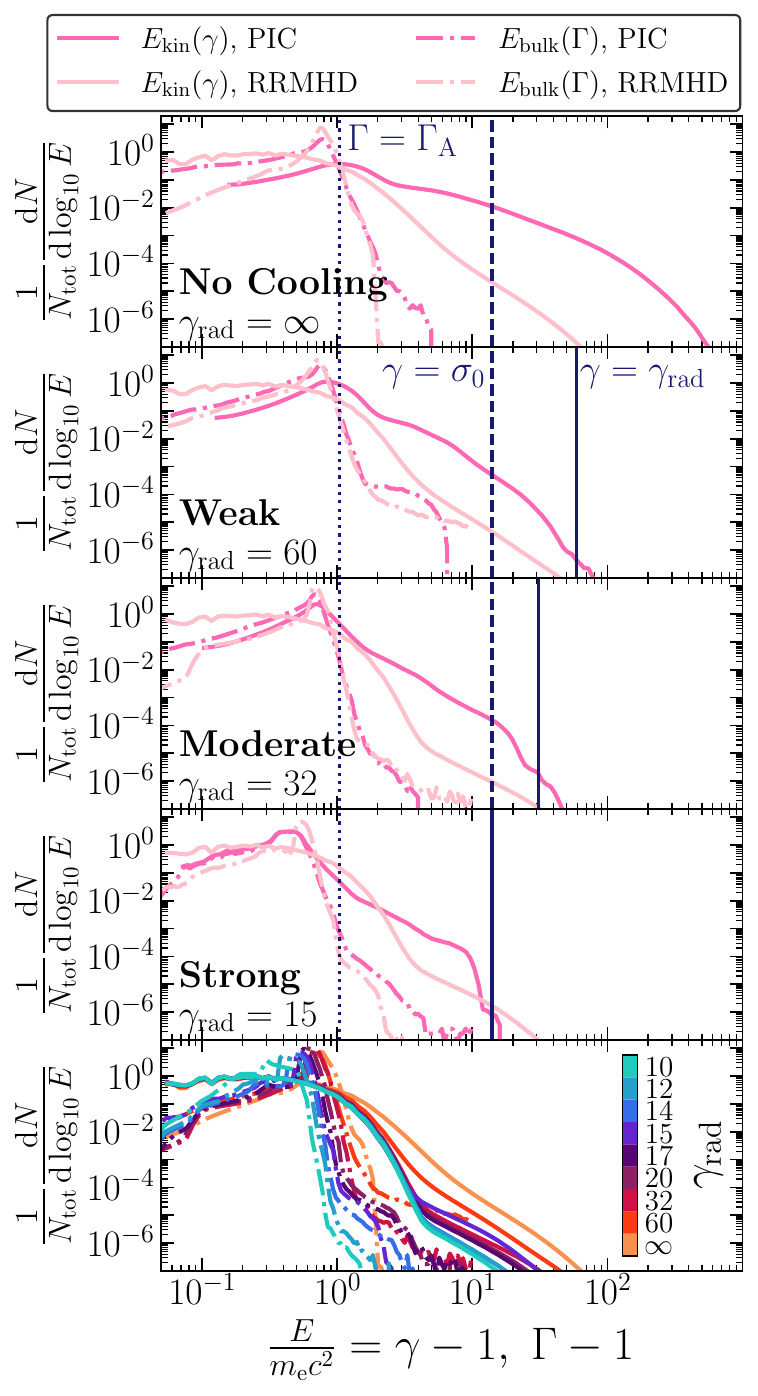}
    \caption{The kinetic energy ($E_\mathrm{kin}(\gamma)$; solid lines) and bulk motion energy ($E_\mathrm{bulk}(\gamma)$; dash-dotted lines) spectra for each cooling regime (top four panels) in PIC (pink) and RRMHD (light pink).  Vertical lines show $\Gamma=\Gamma_\mathrm{A}$ (dotted), $\gamma=\sigma_0$ (dashed) and $\gamma=\Grad$ (solid). The spectra peak at the Alfv\'enic outflow $\Gamma_\mathrm{A}$, and the tails of the energy spectra cool to $\gamma\sim\Grad$. %($\gamma\sim\sigma$ in the case of strong cooling for $\Grad=\sigma=15$).
    All spectra are normalized by $E_\mathrm{tot}$, illustrating the fraction of total energy supplied by bulk motions.  The bulk motion energy spectra match well between RRMHD and PIC for all four cooling regimes; the kinetic energy spectra agree best with smaller $\Grad$ when the high energy tail in PIC is cooled to match the thermal RRMHD distribution. The bottom panel shows the kinetic energy and bulk motion energy spectra for all cooling strengths outlined in Table~\ref{tab:cooling} for RRMHD only, highlighting the work of cooling to bring the distributions to lower energies.
    }
    \label{fig:spec}
\end{figure}

The most quantitative comparison between the two models is in the energy spectra.  We compare both the kinetic energy $\Ek(\gamma)$ spectra and bulk motion energy $\Eb(\Gamma)$ spectra from PIC and RRMHD for each cooling regime in Figure~\ref{fig:spec}.  
The kinetic energy spectra are calculated from the particle energies in PIC ($\Ek=(\gamma-1) mc^2$).  In RRMHD, the kinetic energy spectra are calculated assuming a drifting Maxwellian in each cell, weighted by that cell's density relative to the upstream and cell size relative to the base grid resolution (to account for adaptive mesh refinement), and integrated over the cells in the relevant simulation domain (described below) at each output time step.  Additional details regarding the computation of the kinetic energy spectra in RRMHD are provided in Appendix~\ref{app:MJ}.  The bulk motion energy spectra ($\Eb=(\Gamma-1) mc^2$) are found from the bulk Lorentz factor in each cell. 
%via the bulk three-velocities $\Gamma\equiv\Gamma(v_\Gamma)$; in PIC $v_\Gamma$ is calculated from the magnitude of the current density $|\JJ|=\varrho |\vv_\Gamma|$, where $\varrho$ is the charge density.  

Each spectrum is calculated in cells located in the reconnection layer; for the weak and no cooling regimes (where we expect $\Theta\sim\sigma_0$ in the layer)  this is identified by cells with a temperature above a threshold $\Theta>\Theta_\mathrm{tresh}=3\times10^{-2}$, while for the moderate and strong cases the sheet becomes too cold to be identified by temperate alone, so an additional bulk four-velocity threshold is applied $u_x^2>u_{x,\mathrm{thresh}}^2=0.1$.  The spectra are time-averaged over the steady state of the simulation run from $3.5\leq t/t_c\leq10$.  Finally, each spectrum is normalized by its own $N_\mathrm{tot}$ such that it integrates to 1.%Finally, both the kinetic and bulk motion energy spectra are normalized by the total energy $E_\mathrm{tot}=\int\Ek(\dd N/\dd\log_{10}\Ek)\dd\log_{10}\Ek$, respective to each individual simulation.  

The kinetic energy spectra from RRMHD and PIC exhibit similar peaks in the thermal components, indicating good agreement in temperature in all four cooling regimes between the two models.  The disparity in the tails between PIC and RRMHD -- especially evident without cooling in the top panel -- is explained by the single-fluid collisional nature of RRMHD and its inability to self-consistently capture particle acceleration as in PIC.  The shape of the PIC tail is described by particle acceleration in the diffusion region up to $\gamma\sim\sigma_0$ (dotted vertical blue line in Figure~\ref{fig:spec}), causing a break in the tail, followed by secondary acceleration processes up to $\gamma\sim\Grad$ (solid vertical blue line in Figure~\ref{fig:spec}), where the tail is cut off \citep{Sironi2014}.  The slope after the break steepens with stronger cooling which has a more significant effect on the nonthermal particles in PIC, driving the distribution closer to that in RRMHD.  Thus, stronger cooling shrinks the gap between the spectra from the two models,
%Stronger cooling drives the distribution closer to a thermal state, and thus shrinks the gap between the RRMHD and PIC spectra, 
with both tails fixing to energies described by $\gamma\sim\Grad$ and specifically $\gamma\sim\sigma_0$ in the strongest cooling case%\footnote{Note that while Lorentz factors beyond $\Grad$ are suppressed, they are not unattainable \citep{Sironi2020}.}
.  %The discrepancy between the tails can be eliminated entirely with the inclusion of test particles in MHD \ams{cite/allude to Tanisha (uncooled), Ma\"{e}lle (cooled)}.

The bulk motion energy spectra in PIC and RRMHD show excellent agreement in both the peak and slope for all cooling strengths, and are consistent with previous studies of guide field-mediated reconnection \citep{Gupta2024}.  The peak occurs just below the analytic Alfv\'{e}n Lorentz factor estimate $\Gamma_\mathrm{A}=1/\sqrt{1-\beta_\mathrm{A}^2}$, with $\beta_\mathrm{A}^2=\sigma_0 /\left[1+\sigma_0\left(1+B_g^2/B_0^2\right)\right]$ \citep{Sironi2025} (dotted vertical blue line in Figure~\ref{fig:spec}), indicating the bulk outflow.  %We have over-plotted a Maxwellian distribution of bulk temperature $k_\mathrm{B}T=100\mathrm{\ keV}$, and find such a distribution is well-traced by our bulk spectra\footnote{Comparing with Figure~8 in \ams{lorenzo 2020}, our bulk motions spectra differ slightly more from the Maxwellian due to our guide field being a factor of five stronger.  Note the true test of our cooling model is the comparison to PIC results.}.  
%We have normalized both the bulk motion energy spectrum and the kinetic energy spectrum by the total energy $E_\mathrm{tot}$, and find that \ams{x\%} of the total energy goes into bulk motions.  

RRMHD only recovers the bulk motions energy spectrum with the fast reconnection rate $\brec\sim0.1$, and therefore the charge-starved resistivity model is essential to accurately model the outflow.  The tension of the magnetic field lines that determines the outflow and resulting bulk motions is regulated by the aspect ratio of the diffusion regions, which itself is controlled by the reconnection rate \citepalias{Ripperda2026}. Matching the kinetic rate is therefore not only essential to capture the bulk motion energy spectra accurately, but also directly affects plasmoid dynamics that can result in observable radiative signatures \citep{Philippov2019,Mahlmann2022}. A comparison to the bulk motion energy spectra produced with the $\sim0.01$ reconnection rate with uniform resistivity is provided in Appendix~\ref{app:uni-eta}.

\section{Conclusions}
% \ams{
% \begin{itemize}
%     \item summary/big takeaway: combo of special $\eta$ + cooling gives macro kinetics (rec rate, bulk)
%     \item limitations/future work: test particles, ions, global, thick optical depth, synchrotron, 3D
%     \item first serious attempt at connecting global MHD dynamics with sub-grid scale kinetic physics in extreme radiative 
%     astro environments. additional physics effects (e.g., ions, synchrotron, optically thick) can be build on top of this 
%     framework
% \end{itemize}
% }
This work marks a first step in modelling kinetic aspects of radiative reconnection in RRMHD with applications to extreme radiative astrophysical environments.  We have developed a subgrid model for IC cooling in RRMHD that
reproduces the kinetic physics of radiative reconnection -- the reconnection rate,
plasma temperature, and Compton drag on bulk motions -- in agreement with PIC
simulations of identical setup across all cooling regimes. The model couples a
systematically derived radiative energy and momentum sink, valid in the optically
thin regime, with a localized charge-starved resistivity. Our main findings are:

\begin{enumerate}
    \item \textbf{The charge-starved resistivity is essential to reproduce kinetic properties.} The charge-starved resistivity introduces the effective skin depth which links fluid and kinetic cooling strengths, giving the exact mapping $\Grad^2\to12\Trad^2$ for a thermal distribution. It is also required to reproduce the fast reconnection rate \citepalias{Ripperda2026} and the bulk-motion statistics (Appendix~\ref{app:uni-eta}).

    \item \textbf{Our RRMHD radiative losses combined with the charge-starved resistivity successfully recover the plasma temperature, reconnection rate, and bulk motions distribution from kinetic theory.} Our cooling model is also able to match the peak in the kinetic energy spectrum, indicating agreement in plasma temperature in the reconnection layer; and although the non-thermal tail present in kinetic theory is not achievable through RRMHD, we find better agreement between the PIC and RRMHD kinetic energy spectra with stronger cooling, as the cooling acts to reduce the energies to a more thermal-like distribution. 
\end{enumerate}

Our framework is built to be extended with additional physical effects:
\begin{itemize}
    \item \textbf{Electron-ion plasmas. }Our simulations invoked a pair ($e^\pm$) plasma, and thus our model is missing effects related to multiple species with a mass difference, such as the division of heating from reconnection between ions and electrons.  Recent work has focused on two-temperature MHD \citep{Ressler2015,Liska2022} or electron-ion kinetic simulations \citep{Groselj2026,Mancini2026}, and the inclusion of such a model with an adapted two-temperature version of our radiative cooling could make for more realistic radiative reconnection physics in global simulations of compact objects.   
    \item \textbf{Additional emission processes and gemoetries.} Our cooling sink can be extended to include other types of emission, such as losses from synchrotron or synchrotron self Compton cooling %\ams{(Ma\"elle)}
    , or applied to more complex systems, such as a turbulent flow %\ams{(Maya)}
    , where reconnection layers form dynamically. 
    \item \textbf{Non-thermal particles.} Since RRMHD with charge-starved resistivity captures the magnitude and shape of the non-ideal electric field in the diffusion region \citepalias{Moran2025,Ripperda2026}, we expect that test particles in RRMHD can faithfully reproduce the non-thermal tail of the PIC spectra even in the non-radiative case, which we will explore in future work.
    \item \textbf{Optically thick systems.} Our cooling model only applies in the optically thin regime; to study optically thicker systems -- which may occur in black hole accretion disk coronae, gamma-ray bursts, or magnetar magnetospheres -- the plasma-radiation interaction (i.e., the feedback of a dynamically changing photon field on the plasma) should be taken into account \citep{McKinney2014,Liska2022,Zhang2026,Nagele2026}.  Physical models for pair production and annihilation (e.g., as a source in the continuity equation) or synchrotron self Compton radiation may become important in such applications as well \citep{Beloborodov2017,Nattila2024,Hakobyan2025}. 
    \item \textbf{Global systems.} Particularly in global simulations, a functional form for the energy density $\urad$ evolved in time and space and sourced by e.g., an accretion disk is required, rather than a constant, isotropic photon field as we have assumed here for simplicity. 
\end{itemize}

Our model is designed for application to global GRMHD simulations of luminous systems -- such as in the coronae of accreting black holes, astrophysical jets, or the magnetospheres and winds of compact objects -- where the strength of the cooling can be tuned to a specific environment or regime via the free parameter $\Grad$.  The radiative losses capture IC cooling in diffusion regions as well as Compton drag limiting bulk motions, combined with a charge-starved reconnection rate that sets realistic flaring timescales, altogether supplying physically-motivated cooling to reconnection sites in global simulations. 

\begin{acknowledgments}
The authors thank Braden Seefeldt-Gail, Gibwa Musoke, Ma\"elle Magnan, Charlotte Savelli, Lorenzo Sironi, Andrei Beloborodov, Chris Thompson, Michael Grehan, Sasha Chernoglazov, and Joonas N\"attil\"a for insightful discussions.  

AMS, BR and NM are supported in part by the Natural Sciences \& Engineering Research Council of Canada (NSERC) [funding reference numbers 568580, RGPIN-2023-04844, and RGPIN-2023-04901], and the Canadian Space Agency (23JWGO2A01).  AMS acknowledges the support of NSERC [funding reference numbers 585825-2023, 589240-2024].  Cette recherche a \'et\'e financ\'ee par le Conseil de recherches en sciences naturelles et en g\'enie du Canada (CRSNG), [num\'eros de r\'ef\'erence 568580, RGPIN-2023-04901, 585825-2023, 589240-2024]. AMS was also supported by the Walter C. Sumner Memorial Fellowship and the Mary and Ron Martin Graduate Fellowship in Astrophysics.  AMS, BR, AP, and AL acknowledges support by a grant from the Simons Foundation (MP-SCMPS-00001470). DG~was supported by the Research Foundation--Flanders (FWO) Senior Postdoctoral Fellowship 12B1424N.  AL and AP acknowledge support from
a BSF grant (2022314)

The computational resources and services used in this work were partially provided by facilities supported by the VSC (Flemish Supercomputer Center), funded by the Research Foundation Flanders (FWO) and the Flemish Government -– department EWI, and by Compute Ontario and the Digital Research Alliance of Canada (alliancecan.ca) compute allocation rrg-ripperda.
\end{acknowledgments}

\appendix

\section{Derivation of the energy-momentum sink term\label{app:sink}}
Here we derive for completeness the energy-momentum sinks in Equations~\eqref{eq:S0-1}--\eqref{eq:Si-1} following \citet[hereafter \citetalias{Golan2015}]{Golan2015}.  The sink term for Compton drag acting on a ``blob" with lab-frame bulk velocity $\bg^i$ is found by averaging over the particle and photon energy distribution functions, and is given by Equation (34) in \citetalias{Golan2015}
\begin{equation}
    S^\mu=-\st c^2\int\frac{\dd^3k}{k^0}\int\frac{\dd^3 p}{p^0}\fr\fe p_\nu k^\nu\left[k^\mu+\frac{\left(p_\nu k^\nu\right)p^\mu}{\me^2c^2}\right],
    \label{eq:S-mu}
\end{equation}
where $p^\mu=(\gamma \me c,\gamma \me c \beta^i)$ is the particle four-momentum, $k^\mu=(k,k^i)$ is the photon four-momentum, $\fe\equiv\fe(p^i)$ is the electron distribution function, and $\fr$ is the photon distribution function.  We calculate the source term in the rest (primed) frame of the blob where $f_\mathrm{e}$ is isotropic, and then transform back to the lab (unprimed) frame at the end. 

We begin with the energy component of the sink term $S^{\prime0}$.  Using $p^\prime_{\nu} k^{\prime\nu}=-p^{\prime0}k^{\prime0}+p^\prime_ik^{\prime i}=-p^{\prime0}k^{\prime0}+p^\prime k^\prime\mu^\prime=-p^{\prime0}k^{\prime0}(1-\beta^\prime\mu^\prime)$, since $k^{\prime0}=k^\prime$, and where $\mu^\prime$ is the cosine of the scattering angle, $S^{\prime0}$ becomes
\begin{align}
    S^{\prime0} &= -\st c^2\int\dd^3k^\prime \fr^\prime\int\dd^3p^\prime\fe^\prime
    \frac{p^\prime_\nu k^{\prime\nu}}{p^{\prime0}k^{\prime0}}\
    \left[
        k^{\prime0} + \frac{
        \left(
            p^\prime_\nu k^{\prime\nu}
        \right)p^{\prime0}}{\me^2c^2}
    \right]\\
    &= -\st c^2\int\dd^3k^\prime \fr^\prime\int\dd^3p^\prime\fe^\prime
    \left[
        -k^{\prime0}\left(
            1-\beta^\prime\mu^\prime
        \right)
        +
        \frac{
            {\left(
                p^{\prime0}
            \right)}^2 k^{\prime0}\left(
                1-\beta^\prime\mu^\prime
            \right)^2
        }{
        \me^2 c^2
        } 
    \right]\\
    &=\st c^2\int\dd^3k^\prime\, k^{\prime0} \fr^\prime\int\dd^3p^\prime\fe^\prime
    \left[ 
        \left(
            1-\beta^\prime\mu^\prime
        \right)
        -\gamma^{\prime2}
        \left(
            1-\beta^\prime\mu^\prime
        \right)^2
    \right].
\end{align}
Rewriting $\fe^\prime(p^{\prime i})$ as $\Tilde{\fe^\prime}(p^\prime,\mu^\prime)$, the momentum integrals become $\iiint\dd^3p^\prime\fe^\prime\to\iint\dd\mu^\prime\,\dd p^\prime\,p^{\prime2}\Tilde{\fe^\prime}$, where the factor of $2\pi$ from averaging over $\phi^\prime$ has been absorbed into $\Tilde{\fe^\prime}$, and the energy component of the sink term further simplifies to
\begin{align}
    S^{\prime0} 
    %&= \st\int\dd^3k\, k^0 \fr\int\dd p\,p^2\Tilde\fe\int\dd\mu
    % \left[ 
    %     \left(
    %         1-\beta\mu
    %     \right)
    %     -\gamma^2
    %     \left(
    %         1-\beta\mu
    %     \right)^2
    % \right]\\
    &= \st c^2\int\dd^3k^\prime\, k^{\prime0} \fr^\prime\int\dd p^\prime\,p^{\prime2}\Tilde{\fe^\prime}
    \left[ 
        2 - 2\gamma^{\prime2}
        \left( 
            1+\frac{\beta^{\prime2}}{3}
        \right)
    \right]\\
    &= -2\st c^2\int\dd^3k^\prime\, k^{\prime0} \fr^\prime\int\dd p^\prime\,p^{\prime2}\Tilde{\fe^\prime}
    \left[ 
        \frac{4}{3}\gamma^{\prime2}\beta^{\prime2}
    \right],
\end{align}
where we used $\gamma^{\prime2}-1=\gamma^{\prime2}\beta^{\prime2}$.  

Now, with $\Ne^\prime=\int\dd^3p^\prime\,\fe^\prime=2\int\dd p^\prime\,p^{\prime2}\Tilde{\fe^\prime}$ and the radiation energy-momentum tensor $T^{\prime\mu\nu}_\mathrm{r}=c\int\dd^3k^\prime(k^{\prime\mu} k^{\prime\nu}/k^{\prime0})\fr^\prime$, the energy component of the sink term becomes
\begin{equation}
    S^{\prime0}=-\frac{4}{3}\st c T^{\prime00}_\mathrm{r}\Ne^\prime\langle\gamma^{\prime2}\beta^{\prime2}\rangle.
\end{equation}
For an isotropic radiation field in the lab frame, the radiation energy-momentum tensor is given by 
% \begin{equation}
%     \begin{cases}
%         % \begin{align}
%             T^{00}_\mathrm{r} &=\ \ \urad\\
%             T^{0i}_\mathrm{r} &=\ \ 0\\
%             T^{ij}_\mathrm{r} &=\ \  \frac{\urad}{3}\delta_{ij},
%         % \end{align}
%     \end{cases}
% \end{equation}
\begin{align}
    T^{00}_\mathrm{r}=\urad; &&
            T^{0i}_\mathrm{r}=0; &&
            T^{ij}_\mathrm{r}= \frac{\urad}{3}\delta_{ij};
\end{align}
where $\delta_{ij}$ is the Kronecker delta function.  Then, to find $T^{\prime00}_\mathrm{r}$ in the rest frame, we apply the Lorentz transformation such that $T^{\prime00}_\mathrm{r}=\Lambda_\mu^0\Lambda_\nu^0T^{\mu\nu}_\mathrm{r}=\Gamma^2\urad(1+\bg^2/3)$, where $\Gamma$ is the bulk Lorentz factor and $\bg^i$ is the bulk three-velocity in units of $c$.  This gives the final form of the energy component in the rest frame
\begin{equation}
    % \boxed
    {
        S^{\prime0}=-\frac{4}{3}\st c\Ne^\prime\urad\left\langle\gamma^{\prime2}\beta^{\prime2}\right\rangle\Gamma^2\left[1+\frac{1}{3}\bg^2\right],
    }
    \label{eq:s0-rest}
\end{equation}
as in Equation~(45) from \citetalias{Golan2015}.

Turning now to the momentum component of the sink term, we follow the same steps: 
\begin{equation}
    S^{\prime i} =  -\st c^2\int\dd^3k^\prime \fr^\prime\int\dd^3p^\prime\fe^\prime
    \frac{p^\prime_\nu k^{\prime\nu}}{p^{\prime0}k^{\prime0}}\
    \left[
        k^{\prime i} + \frac{
        \left(
            p^\prime_\nu k^{\prime\nu}
        \right)p^{\prime i}}{\me^2c^2}
    \right]
\end{equation}
Transforming again $\fe^\prime(p^\prime)\to\Tilde{\fe^\prime}(p^\prime,\mu^\prime)$, we average both $k^{\prime i}$ and $\beta^{\prime i}$ over $\phi^\prime$.  Choosing the local $z^\prime$-axis to be in the $\hat{k}^\prime$ direction, 
$\left\langle k^{\prime i}\right\rangle_\phi
=\int\dd\phi^\prime k^{\prime i}
=k^\prime\mu^\prime\hat{k}^\prime=k^{\prime i}$ and $\left\langle\beta^{\prime i}\right\rangle_\phi
=\int\dd\phi^\prime \beta^{\prime i}
%=\int_0^{2\pi}\dd\phi^\prime\beta^\prime(\sin\theta^\prime\cos\phi^\prime\hat{x}^\prime+\sin\theta^\prime\sin\phi^\prime\hat{y}^\prime+\cos\theta^\prime\hat{z}^\prime)
=\beta^\prime\mu^\prime\hat{k}^\prime$, with the $2\pi$ factor again absorbed into $\Tilde{\fe^\prime}$, and so 
\begin{align}
    S^{\prime i} &=\st c^2\int\dd^3k^\prime f^\prime_\mathrm{r}
    \int\dd p^\prime p^{\prime2}\Tilde{\fe^\prime}\int\dd\mu^\prime
    \left(
        1-\beta^\prime\mu^\prime
    \right)
    \left[
        \left\langle
            k^{\prime i}
        \right\rangle_\phi
        - \gamma^{\prime2}k^{\prime}
        \left\langle
            \beta^{\prime i}
        \right\rangle_\phi
        \left(
            1-\beta^\prime\mu^\prime
        \right)
    \right]\\
    &= \st c^2\int\dd^3 k^\prime\fr^\prime\int\dd p^\prime p^{\prime2}\Tilde{\fe^\prime}\int\dd\mu^\prime
    \left(
        1-\beta^\prime\mu^\prime
    \right)
    \left[ 
        k^{\prime i} - \gamma^{\prime2}\beta^\prime\mu^\prime k^{\prime i}
        \left(
        1-\beta^\prime \mu^\prime
        \right)
    \right]\\
    &= \st c^2\int\dd^3 k^\prime k^{\prime i}\fr^\prime\int\dd p^\prime p^{\prime2}\Tilde{\fe^\prime}
    \left( 
        2 + \frac{4}{3}\gamma^{\prime2}\beta^{\prime2}
    \right)\\
    &= \st c\Ne^\prime T^{\prime0i}_\mathrm{r}
    \left\langle 1+ \frac{2}{3}\gamma^{\prime2}\beta^{\prime2} \right\rangle.
\end{align}
After boosting $T^{\prime0i}_\mathrm{r}$ in the same way, we arrive at the final expression for the momentum component of the sink term in the blob's rest frame
\begin{equation}
    S^{\prime i} = -\frac{4}{3}\st c\Ne^\prime\urad\Gamma^2\bg^i
    \left\langle
        1+\frac{2}{3}\gamma^{\prime2}\beta^{\prime2}
    \right\rangle
    \label{eq:Si-rest}
\end{equation}
as in Equation~(46) from \citetalias{Golan2015}. 

We perform one final Lorentz transformation to $S^{\prime\mu}$ to obtain the sink term in the lab frame.  Starting with $S^{\prime0}$, this is simply
\begin{equation}
    S^0=\Lambda^0_0S^{\prime0}+\Lambda^0_i S^{\prime i}=\Gamma S^{\prime0}+\Gamma\bg^iS^{\prime i},
    \label{eq:S0-boost}
\end{equation}
since we are now boosting by $-\bg^i$.  For $S^{\prime i}$, we first apply the boost:
\begin{align}
    S^i 
    &= \Lambda^i_0 S^{\prime0} + \Lambda^i_j S^{\prime j}\\
    &= \Gamma\bg^i S^{\prime0} + S^{\prime j}
    \left( 
        \delta_{ij} + \left(\Gamma-1\right) \frac{\bg^i \bg^j}{\left(\bg\right)^2}
    \right)\\
    &= \Gamma\bg^i S^{\prime0} + S^{\prime i} + \left(\Gamma-1\right) S^{\prime j}\frac{\bg^i\bg^j}{\left(\bg\right)^2}.
\end{align}
Focusing now on the last two terms, we notice $S^{\prime j}\bg^j\bg^i/(\bg)^2=S^{\prime j}\hat{\beta}_\Gamma\hat{\beta}_\Gamma=S^{\prime j}_\parallel$. Then %, since $S^{\prime i}_\perp=S^{\prime i} - S^{\prime i}_\parallel$, 
the above becomes
\begin{align}
    S^i &= \Gamma\bg^i S^{\prime0} + S^{\prime i} + \Gamma S^{\prime i}_\parallel - S^{\prime i}_\parallel\\
    &= \Gamma\bg^i S^{\prime0} + \Gamma S^{\prime i}_\parallel + S^{\prime i}_\perp.
\end{align}
Since the bulk velocity is entirely along $S^{\prime i}_\parallel$, then $S^{\prime i}_\perp=0$ and $S^{\prime i}_\parallel=S^{\prime i}$, and so 
\begin{equation}
    S^i=\Gamma\bg^i S^{\prime0} + \Gamma S^{\prime i}.
    \label{eq:Si-boost}
\end{equation}

Applying the Lorentz transformations from Equations~\eqref{eq:S0-boost} and \eqref{eq:Si-boost} to $S^{\prime0}$ and $S^{\prime i}$ (Equations~\eqref{eq:s0-rest} and \eqref{eq:Si-rest}, respectively) and rewriting the number density $\Ne$ as a mass density $\rho_\mathrm{e}$, we finally arrive at the sink terms in the lab frame shown in Equations~\eqref{eq:S0-1} and~\eqref{eq:Si-1} of the main text:
\begin{align}
    S^0 
    &= -\frac{4}{3}\frac{\st c}{\me}\urad\rho_\mathrm{e}\Gamma^2
    \left(
        \bg^2 + 12\Te^2
        \left(
            1+\bg^2
        \right)
    \right);\\
    S^i 
    &= -\frac{4}{3}\frac{\st c}{\me}\urad\rho_\mathrm{e}\Gamma^2\bg^i
    \left(
        1 + 4\Te^2
        \left(
            5+\bg^2
        \right)
    \right),
\end{align}
where we have used $\left\langle\gamma^{\prime2}\beta^{\prime2}\right\rangle=12\Te^2$ (dropping the prime as temperature is defined in the rest frame).

\section{Resolving the diffusion region in RRMHD\label{app:resolution}}
\begin{figure*}
    \centering
    \includegraphics[width=\linewidth]{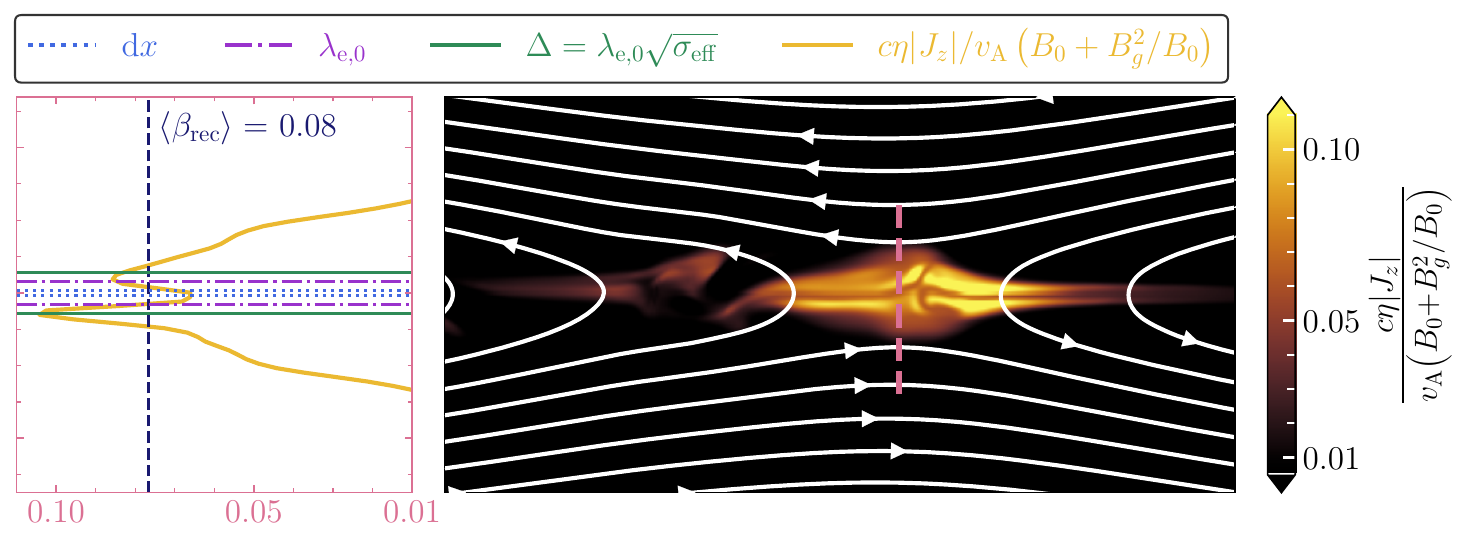}
    \caption{\textbf{Right:} Zoom-in of a diffusion region around an X-point in the RRMHD simulation with $\Grad=15$, showing contours of the out-of-plane vector potential pointing in the direction of the in-plane magnetic field (white lines).  \textbf{Left:} Slice through the diffusion region (indicated by the pink dashed line in the right panel), showing $\langle\brec\rangle\sim c\eta|J_z|/\vA\left(B_0+B_g^2/B_0\right)$.  The separation between two matching horizontal lines illustrates critical length scales -- the cell size $\dd x$ (dotted blue), the effective skin depth $\lambda_\mathrm{e,0}$ (dash-dotted violet), and the sheet half-thickness $\Delta=\lambda_\mathrm{e,0}\sqrt{\sigma_\mathrm{eff}}$ (solid green) -- indicating the resistive scale is well-resolved in our RRMHD simulations.}  
    \label{fig:res}
\end{figure*}
To confirm the resistive scale is resolved even with strong cooling in RRMHD,  we must ensure we are resolving the effective skin depth $\Le$ with a sufficient number of numerical cells.  Based on requirements from PIC simulations \citep{Sironi2025}, as well as RRMHD without cooling \citepalias{Ripperda2026}, we employ 5 cells/$\Le$.  Our base resolution in RRMHD is $256^2$ with eight levels of adaptive mesh refinement, resulting in an effective resolution of $32,768^2$.  The effective skin depth is $\Le\sim m/e/\sqrt{\rho_0}$, with the initial upstream density $\rho_0=B_0^2/\left\{\sigma_0\left[1+\hat{\gamma}\,\Theta_0/\left(\hat{\gamma}+1\right)\right]\right\}$, and where $m/e$ is a free parameter, which should be set such that $\Le\ll L$ \citepalias{Ripperda2026}, similar as in PIC simulations.  With our simulation parameters for initial upstream magnetic field $B_0=1$, magnetization $\sigma_0=15$, dimensionless temperature $\Theta_0=0.01$, adiabatic index $\hat{\gamma}=4/3$, and mass-to-charge ratio $m/e=8\times10^{-5}$, we find $\Le\approx5\dd x\approx0.0003L$, or equivalently $\Delta_0=\lambda_{\mathrm{e},0}\sqrt{\sigma_\mathrm{eff}}\approx0.001L$, where $L$ is the half-length of the simulation domain, $\Delta_0$ is the initial half-thickness of the current sheet, $\lambda_{\mathrm{e},0}$ is the initial upstream cold effective skin depth, and $\sigma_\mathrm{eff}\sim\Gamma_\mathrm{A}\beta_\mathrm{A}$ is the effective magnetization with a guide field \citep{Sironi2025}.

We further illustrate the resolution of our RRMHD simulations in Figure~\ref{fig:res} through a zoom-in of a diffusion region surrounding an X-point.  The right panel shows the effective reconnection electric field $c\eta|J_z|/\vA\left(B_0+B_g^2/B_0\right)\sim\brec$ in the diffusion region.  %This quantity is a proxy for the reconnection rate, as $\brec\sim v_\mathrm{in}/v_\mathrm{out}\sim\left(c\EE\times \BB/B^2\right)/\vA$. 
Here, $|J_z|$ is the magnitude out-of-plane resistive current density:
\begin{equation}
    J_z=\varrho v_z+\frac{\Gamma}{\eta}\left[E_z+\frac{\left(\vv\times\BB\right)_z}{c}-\frac{\left(\EE\cdot\vv\right)v_z}{c^2}\right],
\end{equation}
where $\varrho$ is the charge density \citep{Blackman1993,Komissarov2007}.  In the left panel, we show a vertical slice through the diffusion region and compare it to the time-averaged $\langle\brec\rangle\sim0.08$.  Overtop of the slice of the diffusion region, we show three length scales of interest to the resolution through the distance between pairs of lines: the cell size $\dd x$, the initial effective skin depth $\lambda_\mathrm{e,0}$, and the initial current sheet half-thickness $\Delta_0=\lambda_\mathrm{e,0}\sqrt{\sigma_\mathrm{eff}}$.  The thickness of the diffusion region is larger than $\lambda_{\mathrm{e},0}$ and is shown to approach the expected $\Delta_0$ \citepalias{Ripperda2026}.  The size of $\dd x$ compared to both $\lambda_\mathrm{e,0}$ and $\Delta_0$ confirms the diffusion region is well-resolved for our effective resolution in RRMHD.

\section{Derivation of the Drifting Maxwellian\label{app:MJ}}
Here we derive the drifting Maxwellian calculated in each RRMHD cell to produce the kinetic energy spectrum shown in Figure~\ref{fig:spec}.  We first derive the expression analytically by boosting the local MJ distribution to the lab frame and integrating over the angle between the particle's motion and the bulk motion.  We then show the simplified expression used to calculate the spectrum numerically.
\subsection{Analytical expression}
\citet{Zenitani2015} has shown that the MJ distribution function in terms of four-velocity $\uu$ transforms as
\begin{align}
    f^\prime(\uu^\prime)
    &=f(\uu)= \frac{1}{4\pi\Theta K_2\left(
        \frac{1}{\Theta}
    \right)}
    \exp\left(
        -\frac{\gamma^\prime}{\Theta}
    \right)\\
    &=\frac{1}{4\pi\Theta K_2
    \left(
        \frac{1}{\Theta}
    \right)}
    \exp\left(
        -\frac{\Gamma\gamma\left(
            1-\bg\beta\mu
        \right)}{\Theta}
    \right),
\end{align}
where $\Theta=k_\mathrm{B}T/\me c^2$ is the dimensionless electron temperature, $K_2$ is the modified Bessel function of the second kind, $\Gamma$ is the bulk lorentz factor, $\bg$ is the bulk three-velocity in units of $c$, $\gamma$ is the particle Lorentz factor, $\beta$ is the particle three-velocity in units of $c$, $\mu$ is the cosine of the angle between $\bg$ and $\beta$, and we again indicate the rest frame as primed and the lab frame as unprimed.  Averaging over $\mu$ yields
\begin{align}
    f(\uu) 
    &= \frac{1}{4\pi\Theta K_2\left(
        \frac{1}{\Theta}
    \right)}
    \frac{1}{2}\int_{-1}^1
    \exp\left(
        -\frac{\Gamma\gamma\left(
            1-\bg\beta\mu
        \right)}{\Theta}
    \right)\dd\mu\\
    % &= \frac{1}{4\pi K_2
    % \left(
    %     \frac{1}{\Theta}
    % \right)}
    % \frac{1}{2}
    % \frac{1}{\Gamma\gamma\bg\beta}
    % \left[
    %     \exp\left(
    %         \frac{\Gamma\gamma\bg\beta}{\Theta}
    %     \right) - \exp\left(
    %         -\frac{\Gamma\gamma\bg\beta}{\Theta}
    %     \right)
    % \right]
    % \exp\left(
    %     -\frac{\Gamma\gamma}{\Theta}
    % \right)\\
    % &= \frac{1}{4\pi K_2
    % \left(
    %     \frac{1}{\Theta}
    % \right)}
    % \frac{\sinh\left(
    %     \Gamma\gamma\bg\beta/\Theta
    % \right)}{\Gamma\gamma\bg\beta}
    % \exp\left(
    %     -\frac{\Gamma\gamma}{\Theta}
    % \right)\\
    &= \frac{1}{4\pi K_2
    \left(
        \frac{1}{\Theta}
    \right)}
    \frac{\sinh\left(
        \sqrt{\Gamma^2-1}\sqrt{\gamma^{2}-1}/\Theta
    \right)}{\sqrt{\Gamma^2-1}\sqrt{\gamma^{2}-1}}
    \exp\left(
        -\frac{\Gamma\gamma}{\Theta}
    \right),\\
\end{align}
using $\gamma\beta=\sqrt{\gamma^2-1}$.

Since $f(\uu)\dd^3u=f(\gamma)\dd\gamma\,\dd^3u/\dd\gamma=4\pi\gamma\sqrt{\gamma^2-1}f(\gamma)\dd\gamma$, the final expression for the drifting Maxwellian in terms of $\gamma$ in the lab frame is given by
\begin{equation}
    f(\gamma)\dd\gamma=
    \frac{
        \gamma\sinh\left(
            \sqrt{\Gamma^2-1}\sqrt{\gamma^2-1}/\Theta
        \right)}{
        K_2
        \left(
            \frac{1}{\Theta}
        \right)\sqrt{\Gamma^2-1}
        }
    \exp\left(
        -\frac{\Gamma\gamma}{\Theta}
    \right)
    \dd\gamma.
\end{equation}

\subsection{Numerical calculation}
Equivalently, the boosted MJ distribution can be derived as follows.  We begin with the normalized energy distribution in the fluid rest frame, where the fluid is isotropic:
\begin{equation}
    \dd N=n^\prime f^\prime(\gamma^\prime)\,\dd\gamma^\prime=\frac{n^\prime}{\Theta K_2\left(\frac{1}{\Theta}\right)}\gamma^{\prime2}\beta^\prime(\gamma^{\prime})\exp\left(-\frac{\gamma^\prime}{\Theta}\right)\dd\gamma^\prime.
\end{equation}
Since particle number is invariant,
\begin{equation}
    \dd N
    =\left(
        \frac{\dd N}{\dd\gamma^\prime}
    \right)_\mathrm{rest}\dd\gamma^\prime
    =\left(
        \frac{\dd N}{\dd\gamma}
    \right)_\mathrm{lab}\dd\gamma,
\end{equation}
so,
\begin{equation}
        f(\gamma)=\frac{n^\prime}{n}f^\prime(\gamma^\prime)\frac{\dd\gamma^\prime}{\dd\gamma}=\frac{f^\prime(\gamma^\prime)}{\Gamma}\frac{\dd\gamma^\prime}{\dd\gamma}.
\end{equation}
In the relativisitic limit where $\beta^\prime(\gamma^\prime)\to1$, we have $\dd\gamma^\prime/\dd\gamma\to\gamma^\prime/\gamma$, so that the boosted MJ in the lab frame is given by
\begin{align}
    f(\gamma)\dd\gamma
    &=\frac{\gamma^\prime}{\Gamma\gamma}f^\prime(\gamma^\prime)\dd\gamma^\prime\\
    &=\frac{1}{\Gamma^2(1-\bg\beta^\prime\mu^\prime)}f^\prime(\gamma^\prime)\dd\gamma^\prime\\
    &=\frac{\gamma^{\prime2}\beta^\prime}{\Gamma^2\left(1-\bg\beta^\prime\mu^\prime\right)\Theta K_2\left(\frac{1}{\Theta}\right)}\exp\left(-\frac{\gamma}{\Theta}\right)\dd\gamma^\prime,
\end{align}
which can be numerically binned over a range of rest-frame $\gamma^\prime$ and $\mu^\prime$ to obtain the energy spectrum for a given cell $\Theta$ and lab-frame $\gamma$.

% \section{Plasmoid Statistics?}
\section{Uniform resistivity \label{app:uni-eta}}

\begin{figure}
    \centering
    \includegraphics[width=\linewidth]{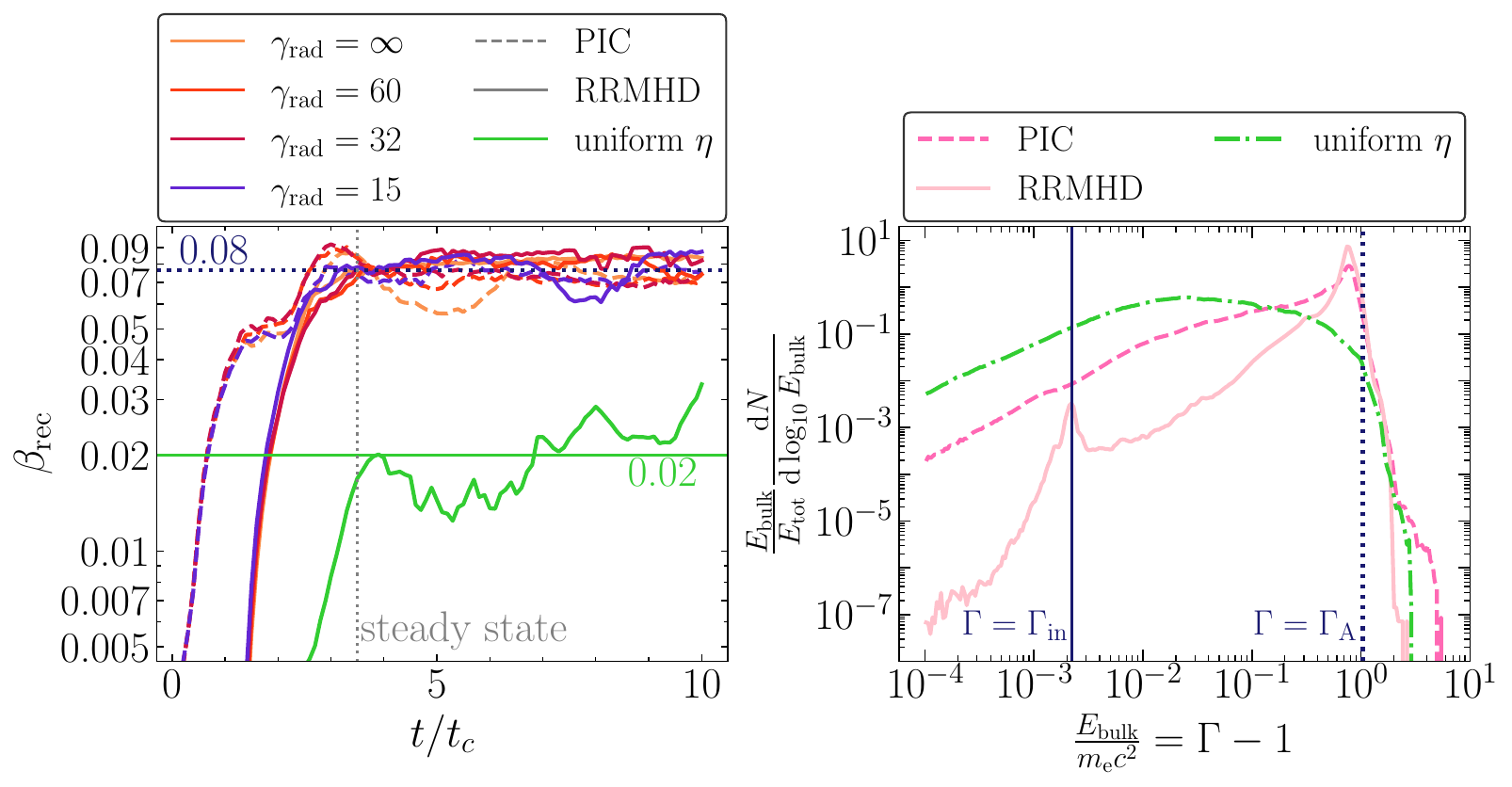}
    \caption{\textbf{Left:} The effect of the charge-starved resistivity on the reconnection rate $\brec$.  Here we plot $\brec$ for the uniform resistivity model with no cooling (green) over the top panel of Figure~\ref{fig:vrec}, comparing it to the fast rate from kinetic simulations or MHD with charge-starved resistivity.  For the kinetic or charge-starved RRMHD, the steady-state time-averaged reconnection rate is $\langle\brec\rangle=0.08$ (dotted navy horizontal line); for the uniform resistivity model, $\langle\brec\rangle=0.02$ (horizontal green line), four times slower.  \textbf{Right:} The bulk motions spectrum $E_\mathrm{bulk}(\Gamma)$ for the no cooling regime in PIC (pink dotted line), charge-starved resistive RRMHD (light pink solid line), and uniform resistivity RRMHD (dash-dotted green line).  The uniform resistivity model cannot reproduce the same peak in the bulk motions spectrum as the kinetic simulation ($\sim\Gamma_\mathrm{A}$), but the charge-starved resistivity model can.  Note the lower energy peak in the charge-starved RRMHD spectrum corresponds to the Lorentz factor of the inflow $\Gamma\sim\Gamma_\mathrm{in}=1/\sqrt{1-\beta_\mathrm{in}^2}\approx2
    \times10^{-3}$.}
    \label{fig:uni}
\end{figure}

The main result of this paper is the reproduction of the PIC macroscopic quantities -- i.e., the reconnection rate, temperature in the reconnection layer, and bulk motions energy spectra -- in RRMHD with IC cooling.  The charge-starved resistivity model is essential in replicating PIC results particularly in the reconnection rate and bulk motion energy distribution.  In Figure~\ref{fig:uni} we compare the reconnection rate (left panel) and bulk motions spectrum (right panel) for an RRMHD simulation with the uniform resistivity prescription to an RRMHD simulation using the charge-starved resistivity and a PIC simulation.  All simulations have cooling turned off to isolate the effect of the $\eta$ prescription and a comparable resolution of $N_x\times N_y=32,768^2$ for both RRMHD simulations and $N_x\times N_y=30,720\times30,000$ in PIC.  With the uniform resistivity model, the reconnection rate is $\sim4$ times slower, and while the maximum value for the bulk spectra remains the same (as it is governed by the Alfv\'enic outflow) the peak is less defined and orders of magnitude less energetic.  The reconnection rate governs the aspect ratio of the diffusion region and thus the field line tension which drives the outflows, and so the slower rate from the uniform resistivity prescription produces a slower average bulk outflow. Therefore, uniform resistivity prescriptions or ideal MHD with a uniform numerical resistivity \citep{Grehan2025} are unable to capture both the kinetic $\brec$ and thus the correct bulk motions, and hence ideal GRMHD \citep[e.g.,][]{Sridhar2025} is not sufficient to model bulk motions due to reconnection in global accretion simulations.

%TC:endignore
%% For this sample we use BibTeX plus aasjournalv7.bst to generate the

\bibliography{noble}{}
\bibliographystyle{aasjournalv7}



\end{document}